\documentclass[Letterpaper]{JHEP3}
\usepackage{epsfig,array,cite,latexsym,amsmath,enumerate}
\usepackage{amssymb}
\usepackage{amsmath}
\usepackage{graphicx}
\usepackage{dcolumn}
\usepackage{bm}
\usepackage{epsfig}
\preprint{INT-PUB 06-06}
\newcommand\beq{\begin{eqnarray}}
\newcommand\eeq{\end{eqnarray}}

\newcommand\bal{ \begin{align}}
\newcommand\eal{\end{align} }

\newcommand\eqn[1]{\label{eq:#1}} 
 
\newcommand\eq[1]{eq.~\eqref{eq:#1}} 
\newcommand\half{{\textstyle{\frac{1}{2}}}} 
\newcommand\fourth{{\textstyle{\frac{1}{4}}}}

\newcommand\bfF{\mathbf{F}}

\newcommand\bfV{\mathbf{V}}

\newcommand\bfZ{\mathbf{Z}}

\newcommand\bfLambda{\boldsymbol{\Lambda}}
 
\newcommand\bfXi{\boldsymbol{\Xi}}

\newcommand\bfPsi{{\mathbf \Psi}}

\newcommand\bfPhi{\boldsymbol{\Phi}}

\newcommand{\CO}{{\cal O}}
\newcommand{\CN}{{\cal N}}

\newcommand{\CQ}{{\cal Q}}

\newcommand{\CL}{{\cal L}}

\newcommand{\bfn}{{\bf n}}

\newcommand{\bfA}{{\bf A}}

\newcommand{\bfe}{{\bf  e}}

\newcommand{\Tr}{{\rm Tr\,}}
\newcommand{\sla}[1]%
        {\kern .25em\raise.18ex\hbox{$/$}\kern-.60em #1}

\newcommand{\mybar}[1]%
        {\kern 0.6pt\overline{\kern -0.6pt#1\kern -0.6pt}\kern 0.6pt}
\newcommand{\dig}{\kern-1.5pt \raisebox{.9ex}{$\cdot$}  \kern1.5pt
  \raisebox{0ex}{${\mathbf\cdot}$}\kern1.5pt \raisebox{-.9ex}{$\cdot$}} 
\newcommand{\digb}{\kern-1.5pt \raisebox{.75ex}{$\cdot$}  \kern1.5pt
  \raisebox{0ex}{${\mathbf\cdot}$}\kern1.5pt \raisebox{-.75ex}{$\cdot$}} 
\newcommand{\digc}{\kern-1.5pt \raisebox{1.05ex}{$\cdot$}  \kern1.5pt
  \raisebox{0ex}{${\mathbf\cdot}$}\kern1.5pt \raisebox{-1.05ex}{$\cdot$}} 
\newcommand{\drawsquare}[2]{\hbox{%
\rule{#2pt}{#1pt}\hskip-#2pt
\rule{#1pt}{#2pt}\hskip-#1pt
\rule[#1pt]{#1pt}{#2pt}}\rule[#1pt]{#2pt}{#2pt}\hskip-#2pt
\rule{#2pt}{#1pt}}
\newcommand{\Yfund}{\raisebox{-.5pt}{\drawsquare{6.5}{0.4}}}
\newcommand{\Ybarfund}{\mybar{\raisebox{-.5pt}{\drawsquare{6.5}{0.4}}}}%
\newbox\pippobox

\title{Lattice formulation of $(2,2)$ supersymmetric gauge theories
  with matter fields}
\author{Michael G. Endres, David B. Kaplan \\ Institute for Nuclear Theory, University of Washington,
  Seattle, WA 98195-1550 \\Email: \email{endres@u.washington.edu},\email{dbkaplan@phys.washington.edu}}
\abstract{
We construct lattice actions for a variety of $(2,2)$ supersymmetric
  gauge theories in two dimensions with matter fields interacting via
  a superpotential.}

\begin{document}




\section{Introduction}
\label{sec:1}
In recent years there has been rapid progress in understanding how to
construct lattice actions for a variety of continuum supersymmetric
theories (see ref.~\cite{Catterall:2005eh} for a recent summary). Supersymmetric gauge
theories are expected to exhibit many
fascinating nonperturbative effects; furthermore, in the limit of large gauge
symmetries, they are related to  quantum gravity and string theory. A
lattice construction of such theories provides a nonperturbative
regulator, and not only establishes that such theories make sense,
but also makes it possible that these theories may eventually be solved
 numerically.
Although attempts to construct supersymmetric lattice theories have
been made for several decades, the new development has been
understanding how to write lattice actions which at finite lattice
spacing possess an exactly realized subset of the 
continuum supersymmetries and have a Lorentz invariant continuum limit.  These exact supersymmetries in many
cases have been shown to constrain relevant operators to the point
that the full supersymmetry of the target theory is attained without a fine tuning.  We will refer to lattices
which possess exact supersymmetries as  ``supersymmetric lattices''.
For alternative approaches
where supersymmetry only emerges in the continuum limit, see \cite{Feo:2002yi,Elliott:2005bd}.

 There have been two distinct
approaches in formulating  supersymmetric lattice actions, recently reviewed in Ref.~\cite{Giedt:2006pd}.  One involves
a Dirac-K\"ahler construction  \cite{Rabin:1981qj,Becher:1982ud} which
associates the Lorentz spinor supercharges with tensors under a diagonal
subgroup of the product of Lorentz and $R$-symmetry
 groups of the
target theory. (An $R$-symmetry is a global symmetry which does not
commute with supersymmetry).  These tensors can then be given a
geometric meaning, with $p$-index tensors being mapped to $p$-cells  on
a lattice.  A lattice action is then constructed from the target
theory which preserves the scalar supercharge even at finite lattice
spacing~\cite{Catterall:2001fr,Catterall:2001wx,Catterall:2003uf,Catterall:2003wd,Sugino:2003yb,Catterall:2004np,Giedt:2005ae,Catterall:2005fd,Catterall:2006jw,Sugino:2006uf}.
This work was foreshadowed by 
an early proposal to use Dirac-K\"ahler fermions in the construction
of a supersymmetric lattice Hamiltonian in one spatial dimension \cite{Elitzur:1982vh}.  
A more ambitious construction which purports to preserve all
supercharges on the lattice  has been proposed 
\cite{Kawamoto:1999zn,Kato:2003ss,D'Adda:2004jb,Kato:2005fj,D'Adda:2005zk},
but  remains controversial \cite{Bruckmann:2006ub}.

The other method for constructing supersymmetric lattices, and the one
employed in this
Letter, is to start with a ``parent theory''--basically the target theory with a parametrically enlarged gauge
 symmetry--and  reduce it to a  zero-dimensional matrix model. One then
 creates a $d$ dimensional lattice with
 $N^d$ sites by modding out a $(Z_N)^d$ symmetry,  where this discrete symmetry is a particular subgroup of
 the gauge, global, and Lorentz symmetries of the parent theory
 \cite{Arkani-Hamed:2001ie,Kaplan:2002wv,Cohen:2003xe,Cohen:2003qw,Giedt:2003xr,Giedt:2003ve,Giedt:2004qs,Kaplan:2005ta}.  
The process of modding out the discrete symmetry  is called an
orbifold projection.  Substituting the projected variables into the
matrix model yields the lattice action.
The continuum limit is  then defined  by expanding the theory about a
point in moduli space that moves out to infinity as $N$ is taken to
infinity, as introduced in the method of deconstruction
\cite{ArkaniHamed:2001ca}.     

Although apparently different, these two approaches to lattice
supersymmetry  yield
similar lattices. The reason for this is that  in the orbifold
approach,  the placement of variables on the lattice is determined by
their charges under a diagonal subgroup
of the  product of the Lorentz and $R$
symmetry groups,  in a manner similar to the 
Dirac-K\"ahler construction \cite{Unsal:2006qp}.

To date, supersymmetric lattices have been
constructed for pure supersymmetric Yang-Mills (SYM) theories, as well
as for two-dimensional Wess-Zumino models.  In this Letter we take the
next step and show how to
write down lattice actions for gauge theories with charged matter
fields interacting via a superpotential.  In
particular, we focus on  two
dimensional gauge theories with four supercharges. 
 These are called $(2,2)$ supersymmetric gauge
theories, and are of particular interest due to their relation to
Calabi-Yau manifolds, as discussed by Witten~\cite{Witten:1993yc}.
Our construction also yields general
insights into the logic of supersymmetric lattices. 

\section{$(2,2)$ SYM}
\label{sec:2}

 We begin with a
brief review of the $(2,2)$ pure Yang-Mills theory.  
The field content of the $(2,2)$ SYM theory is a gauge field, a
two-component Dirac spinor, and a complex scalar $s$, with action
\beq
  \CL &=&  \frac{1}{g_2^2}  \, \Tr
  \Biggl(\bigl\vert D_m s\bigr\vert^2 + \mybar \Psi \, D_m
  \gamma_m 
  \Psi + \fourth v_{mn} v_{mn} \cr &&+\sqrt{2}(\mybar \Psi_L [ s, \Psi_R]
  +\mybar\Psi_R [ s^\dagger, \Psi_L]) + \half
  [s^\dagger,s\,]^2\Biggr)\ ;
  \eqn{targ2}
\eeq
both $\Psi$ and $s$ transform as adjoints under the gauge symmetry.
 The first supersymmetric lattice  for a gauge theory was the
 discretization of the above action
 using the orbifold method \cite{Cohen:2003xe}.  
To construct a lattice for this theory, one begins with a parent theory
which is most conveniently taken to be $\CN=1$ SYM in four dimensions
with gauge group $U(kN^2)$, where the gauge group of the target theory
\eq{targ2} is $U(k)$.  The parent theory possesses
 four supercharges, a gauge field $v_\mu$, and a two
component Weyl
fermion $\lambda$ and its conjugate $\mybar\lambda$,  each variable being a
$kN^2$ dimensional matrix.  When reduced to a matrix model in zero
dimensions, the Euclidean theory has a global symmetry
$G_R=SU(2)_L\times SU(2)_R\times 
U(1)$, where the nonabelian part is inherited from the
four dimensional Lorentz symmetry, and the $U(1)$ is the $R$-symmetry
consisting of a phase rotation of the gaugino.  From the three Cartan
generators $L_3$, $R_3$, $Y$ of $G_R$ we construct two independent
charges $r_{1,2}$ under which the variables of the theory take on
charges $0$ and $\pm 1$ (integer values are required for the lattice
construction, and magnitude no bigger than one to ensure only nearest
neighbor interactions on the lattice).  One can define:
\beq
r_1 = -L_3 + R_3 -Y\ ,\qquad 
r_2 = +L_3 + R_3 -Y\ ,
\eqn{rdef}
\eeq
where $Y$ is $1/2$ times the conventionally normalized $R$-charge
in four dimensions.
By writing $v_\mu$, $\lambda$ and $\mybar\lambda$ as
\beq
v_\mu \mybar\sigma_\mu = \begin{pmatrix}
\mybar z_1 &- z_2 \cr \mybar z_2&  z_1 \end{pmatrix}\ ,\quad
\lambda = \begin{pmatrix}\lambda_1\cr \lambda_2 \end{pmatrix}\ ,\quad
\mybar\lambda =  \begin{pmatrix}\mybar\lambda_1 & \mybar\lambda_{2} \end{pmatrix}\ ,
\eeq
where $\mybar\sigma_\mu = \{1,i\vec\sigma\}$, we arrive at the charge
assignments shown in Table~\ref{tab:rgauge}.  The $N^d$ site lattice is then
constructed by assigning to each variable a position in the unit cell
dictated by its $\vec r = \{r_1,r_2\}$ charges, where $\{0,0\}$
corresponds to a site variable, $\{1,0\}$ corresponds to an oriented
variable on the $x$-link, etc.  Thus from the charges in
Table~\ref{tab:rgauge} we immediately arrive at the lattice structure
shown in Fig.~\ref{twotwo}.

\EPSFIGURE[b!]{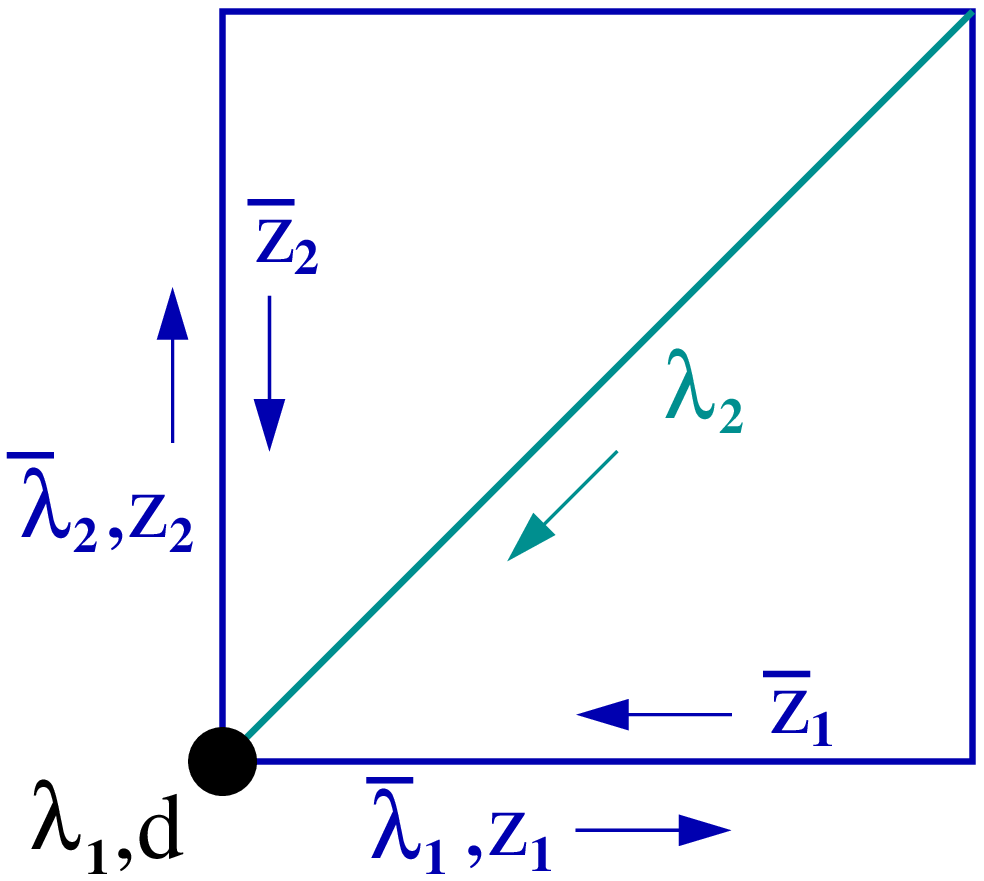,width=2.2in}
{The lattice for pure $(2,2)$ gauge theory, from
 Ref.~\cite{Cohen:2003xe}.  \label{twotwo}
} 

\setlength{\extrarowheight}{5pt}
\TABULAR[t]{c| c c c c | c c c c | c}{
\hline
&$  z_1 $&$ \mybar z_1 $&$ z_2 $&$ \mybar z_2 $&$ \lambda_1 $&$\lambda_2 $&$ \mybar\lambda_1 $&$ \mybar\lambda_2 $& $d$
\\ \hline
$ L_3 $&$ -\half $&$ +\half $&$ +\half $&$ -\half $&$ \,\ 0 $&$ \,\ 0 $&$ -\half $&$ +\half $
&$0$\\
$ R_3 $&$ +\half $&$ -\half $&$ +\half $&$ -\half $&$+\half $&$ -\half $&$ \,\ 0 $&$ \,\ 0 $
&$0$\\
$ Y $ &$ \,\ 0 $&$ \,\ 0 $&$ \,\ 0 $&$ \,\ 0 $&$ +\half $&$ +\half $&$ -\half $&$ -\half $
&$0$\\ \hline
$ r_1 $&$ +1 $&$ -1 $&$ \,\ 0 $&$ \,\ 0 $&$ \,\ 0 $&$ -1 $&$ +1 $&$ \,\ 0 $
&$0$\\
$ r_2 $&$ \,\ 0 $&$ \,\ 0 $&$ +1 $&$ -1 $&$ \,\ 0 $&$ -1 $&$ \,\ 0 $&$ +1 $
&$0$\\ \hline}
{ The $r_{1,2}$ charges of the gauge multiplet. \label{tab:rgauge}}


 The orbifold lattice construction
technique also renders writing down the lattice action a simple
mechanical exercise; here we summarize the results of Ref.~\cite{Cohen:2003xe}.
The lattice variables in Fig.~\ref{twotwo} are $k$ dimensional
matrices, where Greek letters
correspond to Grassmann variables, while Latin letters
are bosons. The lattice action possesses a $U(k)$ gauge symmetry and
single exact supercharge which can be
realized as $Q=\partial/\partial\theta$, where $\theta$ is a Grassmann
coordinate.  
 To make the supersymmetry manifest, the
variables are organized into superfields  as
\beq
\bfZ_{i,\bfn} &=& z_{i,\bfn} + \sqrt{2}\,\theta\,\mybar\lambda_{i,\bfn} \cr
\bfLambda_\bfn&=& \lambda_{1,\bfn} +\theta\,\bigl[\left(z_{i,\bfn}\mybar
  z_{i,\bfn} - \mybar z_{i,\bfn-\bfe_i} z_{i,\bfn-\bfe_i} -i d_\bfn \right)\bigr] \cr
\bfXi_{ij,\bfn}&=& \lambda_{2,\bfn} \epsilon_{ij}  + 2\,\theta \left(\mybar
z_{i,\bfn+\bfe_i}\mybar z_{j,\bfn} - \mybar z_{j,\bfn+\bfe_j}\mybar
z_{i,\bfn}\right)\ ;
\eqn{gmult}
\eeq
a sum over repeated $i$ indices being  implied, where $\bfn$ is a lattice
vector with integer components, and  
$\bfe_i$ is a unit vector in the $i$ direction. The $\mybar z_i$
bosons are supersymmetric singlets.    The lattice 
action may then be written in manifestly supersymmetric form:
\beq
S& =& \frac{1}{g^2} \sum_{\bfn} \int\! d\theta\,
\Tr\biggl[\half\bfLambda_\bfn\partial_\theta \,\bfLambda_\bfn- \bfXi_{ij,\bfn} \bfZ_{i,\bfn}\bfZ_{j,\bfn+\bfe_i}\cr
&& + \bfLambda_\bfn \left(\bfZ_{i,\bfn}\mybar z_{i,\bfn}-\mybar
z_{i,\bfn-\bfe_i} \bfZ_{i,\bfn-\bfe_i}\right)\biggr]\ .
\eeq
 The continuum limit is defined by expanding about the point in moduli
 space $z_i=\mybar z_i = (1/\sqrt{2} a){\bf 1}_k$, where ${\bf 1}_k$
 is the $k$ dimensional unit matrix and $a$ is identified as the
 lattice spacing, and then taking $a\to 0$ with $L=N a$ and $g_2=g a$
 held fixed.
An additional soft supersymmetry breaking mass term
\beq
\delta S =\frac{1}{g^2} \sum_\bfn a^2 \mu^2 \left(\mybar z_{i,\bfn} z_{i,\bfn} - \frac{1}{2 a^2} \right)
\eeq
may be introduced to the action in order to lift the degeneracy of the moduli and fix the
vacuum expectation value of the gauge bosons. The mass parameter $\mu$ is
chosen to scale as $\mu \sim 1/L$ so as to leave physical properties at length scales smaller
than $1/\mu$ unaffected by this modification to the action.
The lattice action has been shown to converge to the $(2,2)$ target
theory \eq{targ2} with the lattice and continuum variables 
related as  $z_i=\frac{1}{\sqrt{2}}\left(1/a + s_i + i v_i\right)$, where 
\beq
s=\frac{s_1+is_2}{\sqrt{2}}\ ,\quad
\Psi = \begin{pmatrix} \lambda_1 \cr
\lambda_2 \end{pmatrix}\ ,\quad
\mybar\Psi = \begin{pmatrix}  \mybar\lambda_1 &  \mybar\lambda_2 \end{pmatrix}
\eeq
in a particular basis for the Dirac $\gamma$ matrices \cite{Cohen:2003xe}.

\section{Adjoint matter}
\label{sec:3}

We now turn to supersymmetric lattices for gauge theories with matter
multiplets, once again employing the orbifold technique. To illustrate 
the general structure of these theories on the lattice, we first
consider as our target theory a $(2,2)$ gauge theory with gauge group
$G=U(k)$ (with $k=1$ a possibility) and $N_f$ flavors of adjoint matter fields.
The parent theory is a four dimensional
$\CN=1$ theory with gauge group $\tilde G = U(kN^2)$ and chiral
superfields $\bfPhi^a$, $a=1,\ldots,N_f$ transforming as adjoints under
$\tilde G$, and a superpotential $W(\bfPhi)$ that preserves the $U(1)$
$R$-symmetry.

The orbifold projection of the matter fields follows a similar path
from that outlined in the previous section for the gauge
multiplet. Each chiral field $ \bfPhi$ from the parent theory
contributes a  boson $ A$,  auxiliary field $F$,  and two component fermion
$\psi_i$;
$\mybar{\bfPhi}$ contributes barred versions of the same. Once
again, the placement of these variables on the lattice is entirely
dictated by their transformation properties under the global
$SU(2)_R\times SU(2)_L \times U(1)$ symmetry of the parent theory,
which we give in Table~\ref{tab:rmatter}. An ambiguity is apparent in
the assignment of the $U(1)$ symmetry to each field, and we have
assigned in the parent theory a  $U(1)$ charge $y$ to our generic
$\bfPhi$. Without a superpotential, there is freedom to assign
to each chiral superfield an independent value for $y$; however, it is
apparent from Table~\ref{tab:rmatter} that to obtain a sensible
lattice with only nearest neighbor interactions ({\it i.e.} all $r_i$
charges equal to $0$ or $\pm 1$), we are constrained
to choose $y=0$ or $y=1$. The result of this choice is shown in
Fig.~\ref{fig:matter};  in fact, we will need both types of
matter multiplets, since the superpotential $W$ must have  net charge
$Y=1$.

\TABULAR[t]{c|cc|cccc| c  c}{
\hline
&$ A $&$ \mybar A $&$ \psi_1 $&$ \psi_2 $&$ \mybar\psi_1 $&$ \mybar\psi_2 $ & $F$ &
$\mybar F$
\\ \hline
$ L_3 $&$ \,\ 0 $&$ \,\ 0 $&$ \,\ 0 $&$ \,\ 0 $&$ -\half $&$ +\half $
&0&0\\
$ R_3 $&$ \,\ 0 $&$ \,\ 0 $&$+\half $&$ -\half $&$ \,\ 0 $&$ \,\ 0 $
&0&0\\
$ Y $&$ +y $&$ -y $&$ y-\half $&$ y-\half $&$ -y+\half $&$ -y+\half $
&$y-1$&$-y+1$\\ \hline
$ r_1 $&$ -y $&$ +y $&$ 1-y $&$ -y $&$ +y $&$ -1+y $
&$y-1$&$-y+1$\\
$ r_2 $&$ -y $&$ +y $&$ 1-y $&$ -y $&$ -1+y $&$ +y $
&$y-1$&$-y+1$\\ \hline}
{ The $r_{1,2}$ charges of the matter multiplet. \label{tab:rmatter}}

We can organize the chiral multiplet $\bfPhi$ of the parent theory for either
case $y=0,1$ into
lattice superfields:
\beq
\bfA_\bfn &=&  A_\bfn + \sqrt{2} \theta \psi_{2,\bfn} \cr
\bfPsi_{\bfn} &=& \psi_{1,\bfn} - \sqrt{2} \theta F_\bfn \cr
\mybar\bfPsi_{i,\bfn} &=& \mybar\psi_{i,\bfn} + 2 \theta
\,\epsilon_{ij}(\mybar A_{\bfn+\bfe_j} 
\mybar z_{j,\bfn+ y\,\bfe_{12}} - \mybar z_{j,\bfn} \mybar A_\bfn)
\cr 
\mybar\bfF_\bfn &=& \mybar{F}_\bfn - 2 \theta
\,\biggl(\mybar A_{\bfn+\bfe_{12}} \lambda_{2,\bfn+ y\,\bfe_{12}} -
\lambda_{2,\bfn}\mybar A_{\bfn}\cr 
&&+\epsilon_{ij}\epsilon_{ik}\left[\mybar\psi_{k,\bfn+\bfe_j}\mybar
  z_{j,\bfn+  y\,\bfe_{12}} - \mybar z_{j,\bfn+\bfe_i}
  \mybar\psi_{k,\bfn}\right]\biggr)
\eqn{mmult}
\eeq
where  $\bfe_{12} =(\bfe_1+\bfe_2)$ and $\mybar A$ is a supersymmetric singlet.
Note the appearance of $\lambda_2$ and $\mybar z_i$ from the gauge
 supermultiplet \eq{gmult}, which implies  nontrivial
 consistency conditions which can be shown to hold.  In the
 Appendix we make contact between the rather unfamiliar multiplet
 structure in \eq{mmult}, and the more familiar chiral superfields
 from $N=1$ supersymmetry in the $3+1$ dimensional continuum.

In terms of the above fields, the orbifold projection of the parent
theory produces the following lattice kinetic Lagrangian for the matter: 
\beq
\CL_{\text{kin}} &=& \frac{1}{g^2}\int\! d\theta\, \Tr \big[\epsilon_{ij}
\mybar\bfPsi_{i,\bfn}\left(\bfZ_{j,\bfn+ y\,\bfe_{12}}\bfA_{\bfn+\bfe_j}-\bfA_\bfn
  \bfZ_{j,\bfn}\right)\cr &&
+\mybar A_\bfn\left(\bfLambda_{\bfn+ y\,\bfe_{12}}\bfA_\bfn -
  \bfA_\bfn\bfLambda_\bfn\right) -\frac{1}{\sqrt{2}} \mybar\bfF_\bfn
\bfPsi_\bfn\big]\ .
\eqn{adjointkin}
\eeq
The superpotential contributions for the theory are 
\beq
\CL_W &=& \frac{1}{g^2} \Tr\biggl[\left( \int\! d\theta\, \frac{1}{\sqrt{2}} \bfPsi^a W_a(\bfA) \right)
+ \mybar \bfF^a \mybar W_a(\mybar A) -
\mybar\bfPsi_1^a\mybar\bfPsi_2^b \mybar W_{ab}(\mybar A)\biggr]
\eqn{superpot}
\eeq
where $W(\bfA)$ is a polynomial in the $\bfA$ fields with $R$-charge
$y=1$ (and $\mybar W(\mybar A)$ is its conjugate), while $W_a=\partial
W/\partial A^a$ and $W_{ab}=\partial^2 
W/\partial A^a\partial A^b$. The space-time dependence
 has been omitted as it is implied by the gauge invariance
 of the Lagrangian; each term in the superpotential should form a closed loop
on the space-time lattice. One can verify by explicit calculation
that the $\theta$ dependence cancels between the second and third
terms after summing over lattice sites, and therefore the action
 is annihilated by $Q=\partial/\partial\theta$ and is supersymmetric. 

\EPSFIGURE[t]{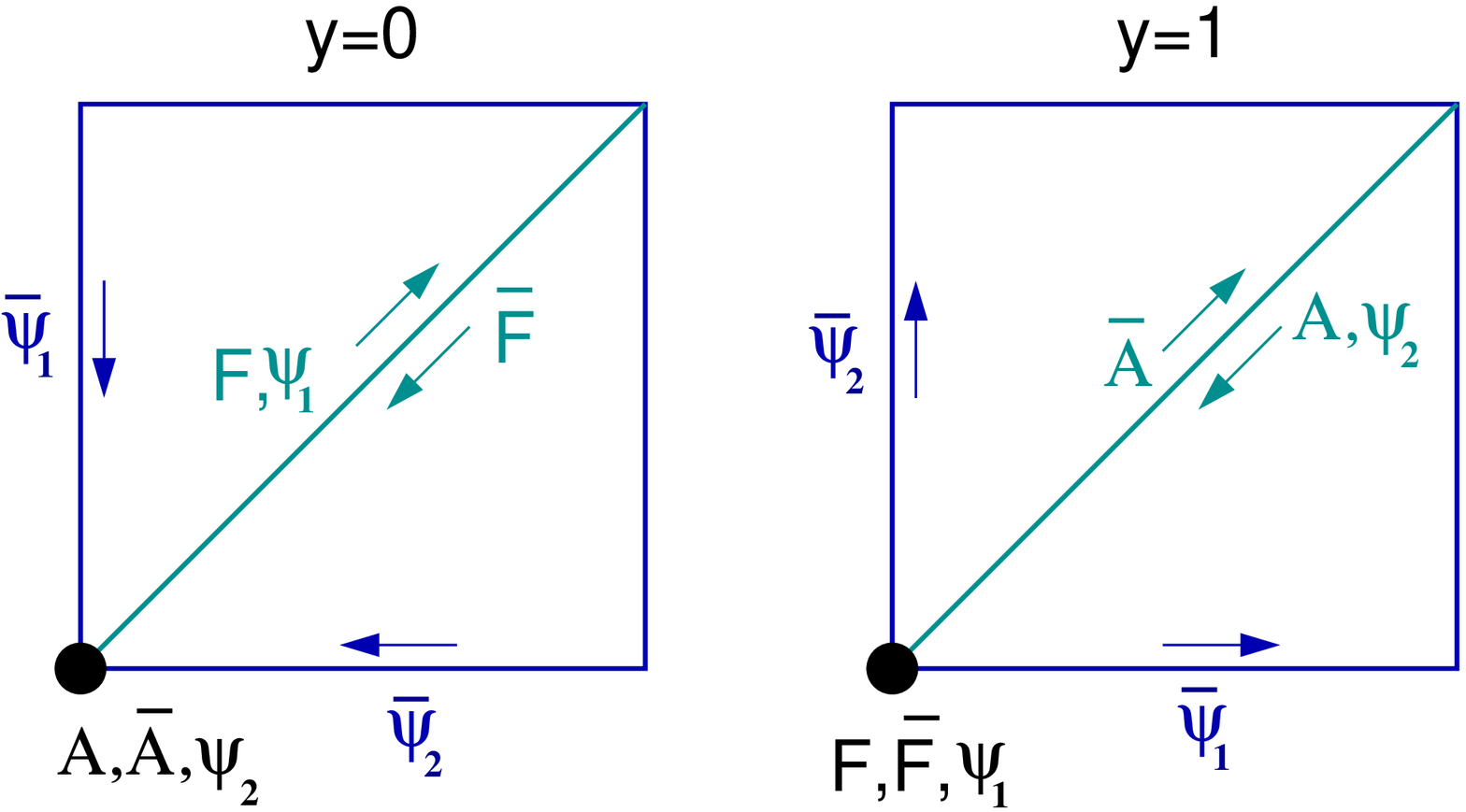, width=4.0in}
{\sl  Placement of the matter variables within the unit cell
 for the two choices $y=0$ and $y=1$ for the $U(1)$ $R$-charge.
\label{fig:matter}}

As an example of how to interpret the above terms, we consider a two flavor
 model ($N_f=2$) and the superpotential $W(\bfPhi)=c\,\Tr\bfPhi^1\bfPhi^2$.
The  superpotential must carry charge $Y=1$, which can be satisfied by
choosing for the superfields $R$-charges $y_1=1$
 and $y_2=0$ for $\bfPhi^1$ and $\bfPhi^2$ respectively.  These charge assignments dictate the lattice
 representation for these superfields, as shown in Fig.~\ref{fig:matter}.
The first term in \eq{superpot}, for example,  is then
\beq
\Tr \bfPsi^a W_a(\bfA) &=&
c\,\Tr\left(\bfPsi^1_{\bfn}\bfA^2_{\bfn} + \bfA^1_{\bfn}\bfPsi^2_{\bfn} \right)
\eeq
which is seen to be gauge invariant since $\{\bfA^1,\bfPsi^1\}$
are $\{-\text{diagonal, site}\}$ variables, while 
$\{\bfA^2,\bfPsi^2\}$ are  $\{\text{site, +diagonal}\}$ variables.

The continuum limit of the above theory is defined as in the previous
section for the pure gauge theory, and the desired $(2,2)$ theory 
with matter results at the classical level.
An  analysis of the continuum limit, including quantum corrections
can be found in the Appendix.
 In the case $k=1$, the
continuum gauge symmetry is $U(1)$ and one obtains 
a theory of neutral matter interacting via a superpotential.

\section{More general  matter multiplets}
\label{sec:4}

More general theories of matter fields interacting via gauge interactions
and a superpotential may be obtained by orbifolding the  parent
theory of \S\ref{sec:3} by some $N$-independent discrete symmetry, before orbifolding
by $Z_N\times Z_N$.  Here we give several examples.

{\it Example 1: $SU(2)\times U(1)$ with charged doublets.} 
Consider the parent theory with a $U(3N^2)$ gauge symmetry,
adjoint superfields $\bfPhi^1$ and $\bfPhi^2$, and the superpotential
 $W(\bfPhi)=c\,\Tr \bfPhi^1\bfPhi^1\bfPhi^2$. Here, we choose $y_1=0$ and $y_2=1$
as R-charges for our superfields.  This theory has
a $\bfPhi^a\to (-1)^a \bfPhi^a$ symmetry, and so we can impose the additional
orbifold condition
 $P\bfPhi^a P = (-1)^a \bfPhi^a$ and  $P \bfV P = \bfV$
where $\bfV$ is the vector supermultiplet of the parent theory and $P$ is
a $ U(3N^2)$ matrix with $\{1,1,-1\}$ along the diagonal, where each
entry is an $N^2$ dimensional unit matrix.  This
projection breaks the $U(3N^2)$ gauge symmetry down to $U(2N^2)\times U(N^2)$,
 under which the projected matter field $\bfPhi^1$ decomposes as
$(\Yfund,\Ybarfund) \oplus (\Ybarfund,\Yfund)$ and $\bfPhi^2$ decomposes as
$({\bf adj},1) \oplus (1,{\bf adj}) $.  We then
orbifold the parent theory by $Z_N\times Z_N$, resulting in a lattice
with an $SU(2)\times U(1)\times U(1)$ gauge theory, with matter
multiplets transforming as
 $3_{0,0} \oplus 2_{\pm 1/2,0} \oplus  1_{0,0} \oplus 1_{0,0}$
 in the continuum limit.
 The doublet couples to both the triplet and one of
the singlets in the superpotential.  Evidently the
second $U(1)$ gauge sector decouples from the theory
since no fields carry that charge.

It is possible to generalize the above construction to fundamental matter
transforming as $\Yfund_{+1} \oplus \mybar\Yfund_{-1}$ under $SU(M) \times U(1)$ gauge
transformations by starting with a $U((M+1)N^2)$ theory broken down to $U(MN^2) \times U(N^2)$.

{\it Example 2:  $U(1)^k$ quiver  with Fayet-Iliopoulos terms.}
A different sort of theory may be obtained by considering a parent 
theory with a $U(kN^2)$ gauge symmetry and a single matter adjoint $\bfPhi$ 
 with a superpotential $W(\bfPhi)=c/k \Tr \bfPhi^k$.  The
initial orbifold condition is  
$\bfV = P \bfV P^\dagger$ and $\bfPhi = \omega P\bfPhi P^\dagger$
on the parent theory, where $\omega = exp(i2\pi/k)$ and $P$ is the diagonal $kN^2$
dimensional ``clock'' matrix
 $\text{diag}\{1,\omega,\omega^2\ldots,\omega^{k-1}\}$, each
entry appearing $N^2$ times.  This projection produces a quiver theory,
breaking the gauge symmetry 
down from $U(kN^2)$ to $U(N^2)^k$, and producing bifundamental matter
fields $\bfPhi^a$, with $a=1,\ldots,k$ transforming as $(\Yfund,\Ybarfund)$ under
$G_a \times G_{a+1}$, where $G_a = U(N^2)$ and $G_{k+1} \equiv G_1$. The superpotential
becomes $W(\bfPhi)=c\Tr \bfPhi^1\cdots\bfPhi^k$. 

One can assign $y=1$ to one of the $k$ matter fields, and $y=0$ to the
others.  A subsequent $Z_N\times Z_N$ projection then produces a lattice
theory with a $U(1)^k$ gauge symmetry, where the descendants of the parent
multiplet $\bfPhi^a$ carry $U(1)$ charges 
 $q_b = (\delta_{ab} -\delta_{a,b-1})$, with $q_{k+1}\equiv q_1$.
One can also add Fayet-Iliopoulos terms to the action given by
$-i\xi\,\int d\theta\, \sum_\bfn\,\Tr \bfLambda^a_\bfn$, as is apparent from
\eq{gmult}.  Such a theory is directly related to Calabi-Yau manifolds,
as discussed in \cite{Witten:1993yc}, and would be interesting to
study numerically.

It should be apparent that although we focused on a $U(1)^k$ quiver,
any $U(p)^q$ quiver can be constructed in a similar manner. We have
not found a way to construct lattices for arbitrary matter representations.

\acknowledgments
We thank Allan Adams and Mithat \"Unsal for enlightening conversations.  This work was
 supported by DOE grants DE-FG02-00ER41132. 

\appendix

\section{Superfield structure}
\label{sec:A1}

The relationship between the lattice superfields defined in \eq{mmult}
and the continuum chiral superfields of the parent
theory can be most easily seen if we turn off the gauge interactions.
Consider the familiar superfield formulation of  $\CN=1$ supersymmetry in four
dimensions. We work in  the superspace coordinate basis  $(y,\,\theta,
\,\mybar\theta)$ from ref.~\cite{Wess:1992cp}, where $\theta$ is a
two-component complex Grassmann 
spinor, and $y_m \equiv (x_m + i \theta
\sigma_m\mybar\theta)$. In this basis the chiral supercharges
$Q_\alpha$ are particularly simple,
\beq
Q_\alpha = \frac{\partial\ }{\partial \theta^\alpha}\ .
\eeq
Furthermore, a  chiral superfield $\bfPhi(y,\theta)$ is independent of
$\mybar\theta$ in this basis, and may be
decomposed as
\beq
\bfPhi &=& A(y) + \sqrt{2}\theta\psi(y) + \theta\theta F(y) \cr
 &=&\bfA(y,\theta^2) + \sqrt{2} \theta^1 \bfPsi(y,\theta^2)\ ,
\eeq
where we follow the spinor notation of \cite{Wess:1992cp}, and
 $\bfA$ and $\bfPsi$ are defined as
\beq
\bfA &=& A(y) + \sqrt{2}\, \theta^2 \psi_2(y)\ ,\cr
\bfPsi &=& \psi_1(y) - \sqrt{2} \,\theta^2 F(y)\ .
\eeq
We see that $\bfA$ and $\bfPsi$ correspond to the first two lattice multiplets in
\eq{mmult}, where the surviving lattice supersymmetry generator is $Q_2 =
\partial/\partial\theta^2$.

The anti-chiral superfield in four dimensions may be written as
$\mybar \bfPhi (\mybar y,\mybar \theta)$.  When this is converted to the  $(y,\,\theta,
\,\mybar\theta)$ basis,  $\mybar \bfPhi$ has the expansion 
\beq
\mybar\bfPhi &=& \mybar A(\mybar y) + \sqrt{2}\mybar \theta\mybar\psi(\mybar y) +
\mybar\theta\mybar\theta \mybar F(\mybar y) \cr &&\cr
&=& 
\mybar A(y) - \sqrt{2} \mybar\theta^j \left[
  \mybar\bfPsi_j(y,\theta^2) + \sqrt{2} \theta^1 \partial_j \mybar
  A(y) \right] \cr 
&&+ \mybar\theta \mybar\theta \left[\mybar\bfF(y,\theta^2) -\sqrt{2}
  \theta^1 \epsilon_{ij} \partial_i \mybar\bfPsi_j(y,\theta^2)
\right]\ , 
\eeq
where
\beq
\mybar\bfPsi_i &=& \mybar\psi_i(y) + \sqrt{2} \theta^2 \epsilon_{ij} \partial_j
\mybar A(y)\ ,\cr
\mybar \bfF &=& \mybar F(y) -\sqrt{2} \theta^2 \partial_j \mybar\psi_j(y)\ .
\eeq
The multiplets $\mybar\bfPsi_i$ and $\mybar \bfF$ are just the continuum
versions of the second two supermultiplets in \eq{mmult}, after
replacing  $\theta\to \theta^2$ and setting to zero the gauge and
gaugino fields. 
Note that the lattice supercharge we have constructed is gauge
invariant, which is why the  gauge  and gaugino fields appear in our
lattice superfields.

With the above packaging, the kinetic energy and superpotential terms for
matter in the lattice theory coincide with those of the parent theory. 
For example, $\CL_{\text{kin}}$ in \eq{adjointkin} takes the familiar form
\beq
\CL_{\text{kin}} &=& \frac{1}{4} \int\! d\theta^2\, d\theta^1\,
d\mybar\theta^1\,  d\mybar\theta^2\, \mybar\bfPhi \bfPhi\ .  
\eeq

\section{Continuum limit and renormalization}
\label{sec:A2}

Radiative corrections and renormalization for the pure (2,2) gauge
theory were considered in Ref.~\cite{Cohen:2003xe}; here we extend
that analysis to include the matter fields interacting through a
superpotential
\beq
W = \Tr \left(\kappa_2^A \bfPhi^A + \kappa_1^{Ab} \bfPhi^A\bfPhi^b +
  \kappa_0^{Abc} \bfPhi^A\bfPhi^b\bfPhi^c \right)
\eeq
where the index $A$ sums over all flavors of $y=1$ matter fields, while $b$, $c$
sum over $y=0$
matter fields (we have normalized the $R$-symmetry such that $W$ has
 $y=1$). 

Induced operators in the Symanzik action take the form
\beq
\delta S_\CO = \frac{1}{g_2^2} \int\! d^2 z\, C_\CO \CO\ ,
\eeq
where $\CO$ is a local operator in the continuum, and $C_\CO$ is a
coefficient depending on the lattice spacing $a$.  The
super-renormalizability of the target theory is most easily accounted
for by defining the scaling dimension of 
$\CO$ according to the  usual conventions of {\it four} dimensional theories:
bosons have mass dimension 1, fermions have mass dimension 3/2, $z$
and $\theta$ have mass dimension $-1$ and $-\half$ respectively.  Then
for an operator $\CO$ of dimension $p$, the coefficient $C_\CO$
induced by radiative corrections takes the form
\beq
C_\CO = a^{p-4}\sum_{\ell=1}^\infty c_\ell (g_2^2 a^2)^\ell\times
F_\ell(\kappa_0,a\kappa_1,a^2\kappa_2)\ ,
\eeq
where $\ell$ corresponds to the number of loops in a perturbative
expansion, and $c_\ell$ is a dimensionless coefficient with only possible
logarithmic dependence on $a$. The functions $F_\ell$ may depend on
both $\kappa_n$ and $\mybar\kappa_n$, but will not diverge as inverse
powers of $a$ as $a\to 0$. 

Induced operators with coefficients which do not vanish as $a\to 0$
will typically spoil the continuum limit of the theory.  However we
see that these could only correspond to $p=2$ at $\ell=1$, $p=1$ at
$\ell=1$, or $p=0$ at $\ell=1,2$.  We can ignore the $p=0$ case, which
corresponds to a cosmological constant and  has no noticeable effects on the
continuum limit.  That leaves us with the only operators to consider
being dimension $p=1$ (scalar
tadpole) or $p=2$ (scalar mass or $F$ tadpole). These operators must
be consistent with the exact symmetries of the lattice: (i) $\CQ=1$
supersymmetry; (ii) the $Z_2$ reflection symmetry about the diagonal
link; (iii) gauge symmetry; (iv) $U(1)$ symmetries.  The latter
include not only the exact $U(1)^3$ global symmetry corresponding to
$r_1$, $r_2$ and $y$, but also the approximate $U(1)^2$ symmetry broken by
the superpotential under which the $\kappa_n$ act as spurions: 
\beq
\bfPhi^a\to e^{i\alpha} \bfPhi^a\ ,\ \ 
\bfPhi^A\to  e^{i\beta}\bfPhi^A
\ ,\ \  \kappa_0 \to e^{-i(2\alpha+\beta)}\kappa_0
\ ,\ \  \kappa_1 \to e^{-i(\alpha+\beta)} \kappa_1
\ ,\ \  \kappa_2 \to e^{-i\beta}\kappa_2\ .\cr
\eqn{u12}\eeq
There may be additional symmetries restricting the form of
counterterms, depending on the form of $W$.

At $p=2$ the operators allowed by symmetry are
\beq
 \int\! d\theta \Tr \bfPsi^A\ ,\qquad \Tr\mybar\bfF^A\ ,\qquad  \Tr \mybar A^a \mybar A^b\ .
\eeq
The second operator does not look supersymmetric, but one can verify
that its $\theta$ component is a total derivative and makes no
contribution to the action.  In each of the above cases it is evident
that the $U(1)^2 $ symmetry of \eq{u12} mandates powers of $a\kappa_1$
and/or $a^2\kappa_2$ in the operator coefficient $C_\CO$, 
 rendering
each of these operators innocuous in the $a\to0$ continuum limit.

At $p=1$ there exists a single operator allowed by the symmetries,
\beq
\Tr\,\mybar A^a\ ,
\eeq
which might  be induced at  one loop with a coefficient
$C_\CO\propto g_2^2 \mybar\kappa_0^{Aab} \kappa_1^{Ab}$ times a possible log. This
contribution can either be calculated and cancelled off by introducing
an explicit tadpole term to the lattice action, or it may be
forbidden by introducing a discrete $\bfPhi^a\to -\bfPhi^a$ 
symmetry, eliminating the $\kappa_1$ coefficient in the superpotential.
In either case, the continuum theory can be attained without any
numerical fine-tuning.

\bibliography{latticeSUSY4}

\providecommand{\href}[2]{#2}\begingroup\raggedright\begin{thebibliography}{10}

\bibitem{Catterall:2005eh}
S.~Catterall, {\it Dirac-kaehler fermions and exact lattice supersymmetry},
  {\em PoS} {\bf LAT2005} (2005) 006,
  [\href{http://xxx.lanl.gov/abs/hep-lat/0509136}{{\tt hep-lat/0509136}}].

\bibitem{Feo:2002yi}
A.~Feo, {\it Supersymmetry on the lattice},  {\em Nucl. Phys. Proc. Suppl.}
  {\bf 119} (2003) 198--209,
  [\href{http://xxx.lanl.gov/abs/hep-lat/0210015}{{\tt hep-lat/0210015}}].

\bibitem{Elliott:2005bd}
J.~W. Elliott and G.~D. Moore, {\it Three dimensional n = 2 supersymmetry on
  the lattice},  {\em PoS} {\bf LAT2005} (2005) 245,
  [\href{http://xxx.lanl.gov/abs/hep-lat/0509032}{{\tt hep-lat/0509032}}].

\bibitem{Giedt:2006pd}
J.~Giedt, {\it Deconstruction and other approaches to supersymmetric lattice
  field theories},  {\em Int. J. Mod. Phys.} {\bf A21} (2006) 3039--3094,
  [\href{http://xxx.lanl.gov/abs/hep-lat/0602007}{{\tt hep-lat/0602007}}].

\bibitem{Rabin:1981qj}
J.~M. Rabin, {\it Homology theory of lattice fermion doubling},  {\em Nucl.
  Phys.} {\bf B201} (1982) 315.

\bibitem{Becher:1982ud}
P.~Becher and H.~Joos, {\it The dirac-kahler equation and fermions on the
  lattice},  {\em Zeit. Phys.} {\bf C15} (1982) 343.

\bibitem{Catterall:2001fr}
S.~Catterall and S.~Karamov, {\it Exact lattice supersymmetry: the
  two-dimensional n = 2 wess-zumino model},  {\em Phys. Rev.} {\bf D65} (2002)
  094501, [\href{http://xxx.lanl.gov/abs/hep-lat/0108024}{{\tt
  hep-lat/0108024}}].

\bibitem{Catterall:2001wx}
S.~Catterall and S.~Karamov, {\it A two-dimensional lattice model with exact
  supersymmetry},  {\em Nucl. Phys. Proc. Suppl.} {\bf 106} (2002) 935--937,
  [\href{http://xxx.lanl.gov/abs/hep-lat/0110071}{{\tt hep-lat/0110071}}].

\bibitem{Catterall:2003uf}
S.~Catterall and S.~Ghadab, {\it Lattice sigma models with exact
  supersymmetry},  {\em JHEP} {\bf 05} (2004) 044,
  [\href{http://xxx.lanl.gov/abs/hep-lat/0311042}{{\tt hep-lat/0311042}}].

\bibitem{Catterall:2003wd}
S.~Catterall, {\it Lattice supersymmetry and topological field theory},  {\em
  JHEP} {\bf 05} (2003) 038,
  [\href{http://xxx.lanl.gov/abs/hep-lat/0301028}{{\tt hep-lat/0301028}}].

\bibitem{Sugino:2003yb}
F.~Sugino, {\it A lattice formulation of super yang-mills theories with exact
  supersymmetry},  {\em JHEP} {\bf 01} (2004) 015,
  [\href{http://xxx.lanl.gov/abs/hep-lat/0311021}{{\tt hep-lat/0311021}}].

\bibitem{Catterall:2004np}
S.~Catterall, {\it A geometrical approach to n = 2 super yang-mills theory on
  the two dimensional lattice},  {\em JHEP} {\bf 11} (2004) 006,
  [\href{http://xxx.lanl.gov/abs/hep-lat/0410052}{{\tt hep-lat/0410052}}].

\bibitem{Giedt:2005ae}
J.~Giedt, {\it R-symmetry in the q-exact (2,2) 2d lattice wess-zumino model},
  {\em Nucl. Phys.} {\bf B726} (2005) 210--232,
  [\href{http://xxx.lanl.gov/abs/hep-lat/0507016}{{\tt hep-lat/0507016}}].

\bibitem{Catterall:2005fd}
S.~Catterall, {\it Lattice formulation of n = 4 super yang-mills theory},  {\em
  JHEP} {\bf 06} (2005) 027,
  [\href{http://xxx.lanl.gov/abs/hep-lat/0503036}{{\tt hep-lat/0503036}}].

\bibitem{Catterall:2006jw}
S.~Catterall, {\it Simulations of n = 2 super yang-mills theory in two
  dimensions},  {\em JHEP} {\bf 03} (2006) 032,
  [\href{http://xxx.lanl.gov/abs/hep-lat/0602004}{{\tt hep-lat/0602004}}].

\bibitem{Sugino:2006uf}
F.~Sugino, {\it Two-dimensional compact n = (2,2) lattice super yang-mills
  theory with exact supersymmetry},  {\em Phys. Lett.} {\bf B635} (2006)
  218--224, [\href{http://xxx.lanl.gov/abs/hep-lat/0601024}{{\tt
  hep-lat/0601024}}].

\bibitem{Elitzur:1982vh}
S.~Elitzur, E.~Rabinovici, and A.~Schwimmer, {\it Supersymmetric models on the
  lattice},  {\em Phys. Lett.} {\bf B119} (1982) 165.

\bibitem{Kawamoto:1999zn}
N.~Kawamoto and T.~Tsukioka, {\it N = 2 supersymmetric model with dirac-kaehler
  fermions from generalized gauge theory in two dimensions},  {\em Phys. Rev.}
  {\bf D61} (2000) 105009, [\href{http://xxx.lanl.gov/abs/hep-th/9905222}{{\tt
  hep-th/9905222}}].

\bibitem{Kato:2003ss}
J.~Kato, N.~Kawamoto, and Y.~Uchida, {\it Twisted superspace for n = d = 2
  super bf and yang-mills with dirac-kaehler fermion mechanism},  {\em Int. J.
  Mod. Phys.} {\bf A19} (2004) 2149--2182,
  [\href{http://xxx.lanl.gov/abs/hep-th/0310242}{{\tt hep-th/0310242}}].

\bibitem{D'Adda:2004jb}
A.~D'Adda, I.~Kanamori, N.~Kawamoto, and K.~Nagata, {\it Twisted superspace on
  a lattice},  {\em Nucl. Phys.} {\bf B707} (2005) 100--144,
  [\href{http://xxx.lanl.gov/abs/hep-lat/0406029}{{\tt hep-lat/0406029}}].

\bibitem{Kato:2005fj}
J.~Kato, N.~Kawamoto, and A.~Miyake, {\it N = 4 twisted superspace from
  dirac-kaehler twist and off- shell susy invariant actions in four
  dimensions},  {\em Nucl. Phys.} {\bf B721} (2005) 229--286,
  [\href{http://xxx.lanl.gov/abs/hep-th/0502119}{{\tt hep-th/0502119}}].

\bibitem{D'Adda:2005zk}
A.~D'Adda, I.~Kanamori, N.~Kawamoto, and K.~Nagata, {\it Exact extended
  supersymmetry on a lattice: Twisted n = 2 super yang-mills in two
  dimensions},  {\em Phys. Lett.} {\bf B633} (2006) 645--652,
  [\href{http://xxx.lanl.gov/abs/hep-lat/0507029}{{\tt hep-lat/0507029}}].

\bibitem{Bruckmann:2006ub}
F.~Bruckmann and M.~de~Kok, {\it Noncommutativity approach to supersymmetry on
  the lattice: Susy quantum mechanics and an inconsistency},  {\em Phys. Rev.}
  {\bf D73} (2006) 074511, [\href{http://xxx.lanl.gov/abs/hep-lat/0603003}{{\tt
  hep-lat/0603003}}].

\bibitem{Arkani-Hamed:2001ie}
N.~Arkani-Hamed, A.~G. Cohen, D.~B. Kaplan, A.~Karch, and L.~Motl, {\it
  Deconstructing (2,0) and little string theories},  {\em JHEP} {\bf 01} (2003)
  083, [\href{http://xxx.lanl.gov/abs/hep-th/0110146}{{\tt hep-th/0110146}}].

\bibitem{Kaplan:2002wv}
D.~B. Kaplan, E.~Katz, and M.~Unsal, {\it Supersymmetry on a spatial lattice},
  {\em JHEP} {\bf 05} (2003) 037,
  [\href{http://xxx.lanl.gov/abs/hep-lat/0206019}{{\tt hep-lat/0206019}}].

\bibitem{Cohen:2003xe}
A.~G. Cohen, D.~B. Kaplan, E.~Katz, and M.~Unsal, {\it Supersymmetry on a
  euclidean spacetime lattice. i: A target theory with four supercharges},
  {\em JHEP} {\bf 08} (2003) 024,
  [\href{http://xxx.lanl.gov/abs/hep-lat/0302017}{{\tt hep-lat/0302017}}].

\bibitem{Cohen:2003qw}
A.~G. Cohen, D.~B. Kaplan, E.~Katz, and M.~Unsal, {\it Supersymmetry on a
  euclidean spacetime lattice. ii: Target theories with eight supercharges},
  {\em JHEP} {\bf 12} (2003) 031,
  [\href{http://xxx.lanl.gov/abs/hep-lat/0307012}{{\tt hep-lat/0307012}}].

\bibitem{Giedt:2003xr}
J.~Giedt, E.~Poppitz, and M.~Rozali, {\it Deconstruction, lattice
  supersymmetry, anomalies and branes},  {\em JHEP} {\bf 03} (2003) 035,
  [\href{http://xxx.lanl.gov/abs/hep-th/0301048}{{\tt hep-th/0301048}}].

\bibitem{Giedt:2003ve}
J.~Giedt, {\it Non-positive fermion determinants in lattice supersymmetry},
  {\em Nucl. Phys.} {\bf B668} (2003) 138--150,
  [\href{http://xxx.lanl.gov/abs/hep-lat/0304006}{{\tt hep-lat/0304006}}].

\bibitem{Giedt:2004qs}
J.~Giedt and E.~Poppitz, {\it Lattice supersymmetry, superfields and
  renormalization},  {\em JHEP} {\bf 09} (2004) 029,
  [\href{http://xxx.lanl.gov/abs/hep-th/0407135}{{\tt hep-th/0407135}}].

\bibitem{Kaplan:2005ta}
D.~B. Kaplan and M.~Unsal, {\it A euclidean lattice construction of
  supersymmetric yang- mills theories with sixteen supercharges},  {\em JHEP}
  {\bf 09} (2005) 042, [\href{http://xxx.lanl.gov/abs/hep-lat/0503039}{{\tt
  hep-lat/0503039}}].

\bibitem{ArkaniHamed:2001ca}
N.~Arkani-Hamed, A.~G. Cohen, and H.~Georgi, {\it (de)constructing dimensions},
   {\em Phys. Rev. Lett.} {\bf 86} (2001) 4757--4761,
  [\href{http://xxx.lanl.gov/abs/http://arXiv.org/abs/hep-th/0104005}{{\tt
  http://arXiv.org/abs/hep-th/0104005}}].

\bibitem{Unsal:2006qp}
M.~Unsal, {\it Twisted supersymmetric gauge theories and orbifold lattices},
  \href{http://xxx.lanl.gov/abs/hep-th/0603046}{{\tt hep-th/0603046}}.

\bibitem{Witten:1993yc}
E.~Witten, {\it Phases of n = 2 theories in two dimensions},  {\em Nucl. Phys.}
  {\bf B403} (1993) 159--222,
  [\href{http://xxx.lanl.gov/abs/hep-th/9301042}{{\tt hep-th/9301042}}].

\bibitem{Wess:1992cp}
J.~Wess and J.~Bagger, {\it Supersymmetry and supergravity}, . Princeton, USA:
  Univ. Pr. (1992) 259 p.

\end{thebibliography}\endgroup
\bibliographystyle{JHEP} 
\end{document}